# Measurement of static light scattering of bio-particles

**Sanchita Roy[a*], Aranya Bhuti Bhattacherjee[a,b], Farhana Hussain[a], Nilakshi Barua[c] and Gazi Ameen Ahmed[d]**

[a]Department of Physics, School of Applied Sciences, University of Science and Technology Meghalaya, District- Ri Bhoi, 793101, Meghalaya, India.
[b]Department of Physics, Birla Institute of Technology and Science, Pilani, Hyderabad Campus, Telangana 500078, India
[c]Department of Microbiology, Faculty of Medicine, The Chinese University of Hong Kong, Prince of Wales Hospital, Shatin, NT, Hong Kong.
[d]Optoelectronics and Photonics Laboratory, Department of Physics, Tezpur University, Napaam, Tezpur- 784028, Assam, India.

## ABSTRACT

Light scattering by small particles is one of the most prevailing and non-invasive technique for examining the properties of particulate systems chiefly of biological origin. The sub-micron particles including the bio-particles were so chosen because of their importance in biology and biomedical sciences. Light scattering investigation from homogenous, pathogenic *Pseudomonas aeruginosa* and non-pathogenic *Mycobacterium smegmatis* was carried out at two different wavelength of incident light, by using an original designed and fabricated polar and azimuth-dependent light scattering setup. Theoretical scattering plots were generated by using T-matrix approach to validate our analyses. Simulations of light scattering of these particles were also carried out using a novel Monte- Carlo simulation technique. The relation between experimental, theoretical and simulated result is presented in this paper. We have studied *Pseudomonas aeruginosa* and *Mycobacterium smegmatis*, and attempted for prospect of observing its morphological property by using light scattering tool.

## 1. Introduction

Light scattering is a vital tool for optical diagnostics of particulate systems. It has several applications in particle characterization and remote sensing of micron and sub-micron particles in the form of aerosols, interplanetary dust, nanoparticles, bacteria, biological cells etc.[1-4]. It is often productive approach to discuss elastic light scattering in terms of Stokes vector and Mueller matrices [4-8]. Variety of laboratory measurements, both *in-situ* and *ex- situ* of visible and infra-red light scattering were performed to determine the phase function and extinction, polarization and depolarization characteristics of natural and artificially generated cirrus cloud crystals [9,10], scattering matrix elements of hydrosols [11,12], backscattering Mueller matrix for atmospheric aerosols [13,14], etc. Still, the complete structure, size, composition and optical parameters of thousands of aerosols and other particulate matters including bio-particles are yet to be done. Among various applications of scattering techniques, biomedical applications find great significance nowadays [14,15]. Light scattering may be able to provide real-time technology to detect and classify different biological species having a definite morphology [15,16]. In the present work, we carried out size quantification of *Pseudomonas aeruginosa* and *Mycobacterium smegmatis* using a novel light scattering instrument. It has been reported earlier that individual *Escherichia coli* cells were studied from light scattering with the scanning flow cytometer by different group of researchers [17-19], however light scattering studies using a setup of our kind is unique and significant [20,21]. This is because our setup includes consideration of both polar and azimuthal dependency of scattered signal. We know that light scattering from single particle or collection of particles is very sensitive to their morphology and optical properties, therefore light scattering can be a sensitive tool for differentiating between cell types, or to distinguish between healthy and malignant cells or to probe changes in cells resulting from stimulus [22-24].

Again, it is important to mention that the studies of organic molecules may also be utilized in interpretation of data that is obtained from Astrophysical studies on the presence of organic molecules in stellar and interstellar medium [ 16,25,26]. In our present work, we found both the samples showing azimuthal dependency. Yet again, experimental light scattering technique alone may not be sufficient to provide complete information about scattering properties of some sub-micron particles especially particles of biological origin.

This calls for need to make use of theoretical approach and computer simulation based on the established theories as an additional tool for typifying such particles.

## 2. Experimental details

**2.1 Light scattering setup:** An original designed and fabricated laser based light scattering system was used to perform scattering experiments [20]. The first two significant elements ($S_{11}$ and $S_{12}$) of the Mueller matrix were measured as a function of the scattering angle $\theta$ corresponding to azimuth angle $\phi$ by using the set-up [20,21]. The elements of the Mueller scattering matrix at any angle represent the effect of scatterer on the intensity and the polarization state of the incident wave [1-4]. These elements contain information about the size, shape, and refractive index etc. of the scattering sample. By measuring the first element of the Mueller matrix, we have determined the volume scattering function for incident light. We used He-Ne lasers as light source, operating at wavelengths 543.5 nm and 632.8 nm respectively for experimental investigations. The laser light was scattered by an ensemble of randomly oriented bioparticles in the scattering volume.

Samples of necessarily low concentration were used to avoid multiple scattering. We normalized all measured scattered intensity profiles to 1 at 10°. The experimental errors are indicated by error bars. It is important to mention that natural particles are often complex (in terms of size, shape, size distribution, refractive index etc.) and show discrepancies between the theoretical and experimentally measured results. Thus, it becomes difficult to characterize and model such particles. In order to overcome such discrepancies and to improve upon the performance, usefulness and quality of light scattering theories, we need to collect sufficient scattering data. Despite advanced numerical techniques being available, the results of such techniques do not completely fit with the results of laboratory and *in-situ* experiments, especially for particles with complicated morphology. This is because of the limitations of such theories when applied to complex particles and to some extent because of the imprecisions that may be present in the experimental setup. Combined efforts on improving and developing computation, and experimentation can produce better results.

**2.2 Sample preparation:**

We made an attempt to find the morphological properties of particular type of bacterial strains namely the *Pseudomonas aeruginosa and Mycobacterium smegmatis*. *Pseudomonas*

*aeruginosa is* commonly found in soil and water and is a free-living bacterium. *P. aeruginosa* is a gram-negative rod-shaped bacterium whose length ranges from 1.5 to 3.0 micrometers and diameter ranges from 0.5 to 0.8 micrometers. *P. aeruginosa* is also called an "opportunistic" human pathogen because it hardly infects healthy individuals but effects patients. Unfortunately, in hospitalized patients, *Pseudomonas* infections are becoming more difficult to treat because it has antibiotic resistance. National Nosocomial Infections Surveillance system from 1986–2003 reported the overall infections caused by *P. aeruginosa* to be stable during 1986–2003, however, an alarming rise in the resistant isolates has been reported in 2003 compared to 1998 through 2002 [28]. Treatment of *P. aeruginosa* infections can be difficult due to its natural resistance to antibiotics. Thus such organisms hold pronounced importance in biomedical sciences [28, 29] and that a proper characterization by non-invasive technique (like light scattering technique) of such drug-resistant bacteria will prove to be a breakthrough in biomedical sciences.

*M. smegmatis is* a gram positive bacterium and it is 3.0 to 5.0 µm long with a bacillus shape [30]. *M. smegmatis* is commonly used in work as the mycobacterium species grows very fast and is non-pathogenic. It holds its importance because it is considered as a model scheme of *Mycobacterium tuberculosis* as it possesses all the characteristics of *M. tuberculosis* except its pathogenic behavior. The direct study of *M. tuberculosis* is vital to understanding its pathogenesis. However, use of this pathogen in the laboratory requires dedicated biosafety safety level 3 laboratory conditions as it carries the risk of accidental exposure [31]. Mycobacterial models have significantly contributed to understanding *M. tuberculosis,* however, the methods were by employment of bio-chemical processes mainly. [31,32]

*M. smegmatis* is also useful for the research analysis of other Mycobacteria species in laboratory experiments [31]. This is because it is a simple model which is easy to work with and it grows very fast, requires only a biosafety level 1 laboratory [ 31]. The ease of time and substantial arrangement needed to work with pathogenic species encouraged researchers to use *M. smegmatis* as a model for mycobacterial species. Moreover, this species shares more than 2000 homologs with *M. tuberculosis* and shares the same uncommon cell wall structure of *M. tuberculosis* and other mycobacterial species. These properties make it a very attractive model organism for *M. tuberculosis* and other mycobacterial pathogens [31]

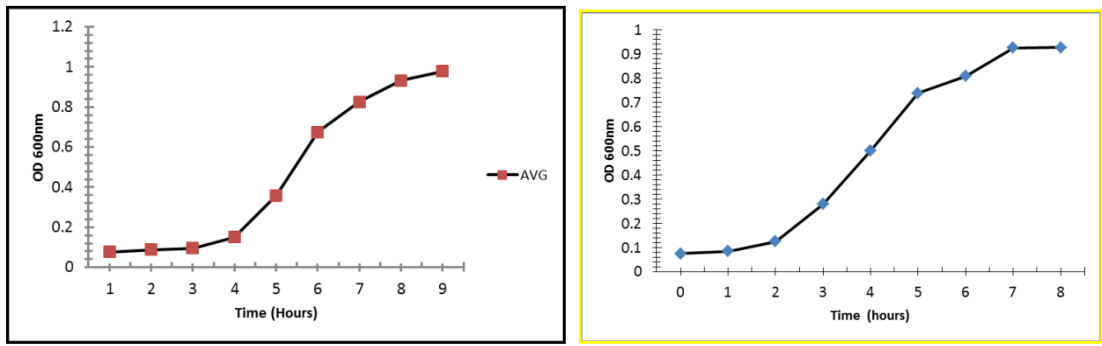

**Figure 1: (a)** Optical density of *P. aureginosa* at 600nm;
**(b)** Optical density of *M. smegmatis* at 600nm

**2.3 Scanning Electron Microscopic investigation (SEM):** The samples were examined under the Scanning Electron Microscope. The dried samples were coated with 10-15 nm thickness of platinum using a JEOL 1600 Auto Fine Coater. The samples were then examined under the Scanning Electron Microscope with an accelerating voltage of 10-15 KV. Scanning Electron Microscopy was performed to see the morphology of *P. aeruginosa* and strain *M. smegmatis* bio-particles. The size distribution profile of these particles was calculated from SEM image. The SEM image of *P. aeruginosa* ensured a rod-shaped structure having an average diameter of 0.65 μm. The size distribution was found to be nearly Gaussian having random orientation of the particles. The SEM image of *M. smegmatis* revealed that the particles were bacillus-shaped having an average diameter of 0.34 μm and with a broad Gaussian size distribution.

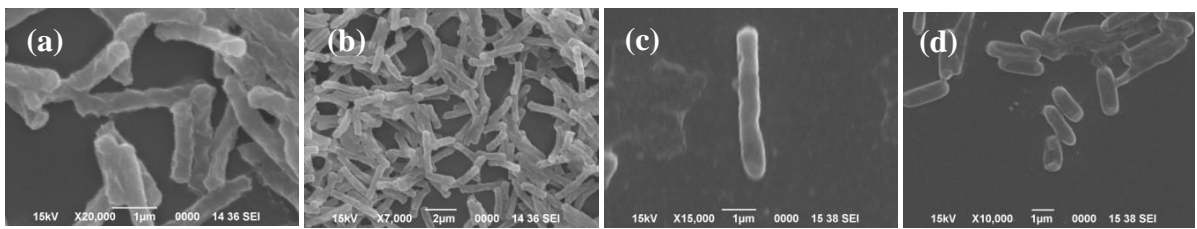

**Figure 2: (a)** SEM of *M. smegmatis* at magnification of X20,000; **(b)** SEM of *M. smegmatis* at magnification of X7,000 **; (c)** SEM of *P. aeruginosa* at magnification of X15,000**;
(d)** SEM of *P. aeruginosa* at magnification of X10,000.

## 3. Data analysis and results

**3.1 Light scattering measurements:** For every light scattering investigation, the scattering detector was first placed at an angle $\theta = 0°$. Then the measurements of scattered light signals were recorded from an angle of 10° to 170°. The scattering measurements were carried out at an incident wavelength of 543.5 nm and 632.8 nm respectively for all the investigations. The scattered light intensity that corresponded to the $S_{11}$ element of the scattering matrix was detected by the detector present in the light scattering setup. Size quantification of such bio- particles could be done by determining the $S_{11}$ element of Mueller matrix. Thus, we could measure the volume scattering function *β(θ)*.

In our present work, we found both the samples showing azimuthal dependency which finally rule out the established impression that randomly oriented particles of any geometry are azimuthally independent. Though both the bacterial samples are rod-like structured, however, each one shows a distinct characteristic profile with azimuthal dependency. The particles possibly took some orientation and alignment which might have created the azimuthal dependency. Thus, we cannot ignore the azimuthal dependency for bioparticles while using light scattering technique which was employed to characterize such particles. On comparison of the theoretical, experimental and simulated results we found discrepancies in correspondence with the profiles at some angles. This may be attributed to the non-inclusion of size distribution function while generating the T-matrix plot. However, while generating the simulated light scattering results, it was done by using a novel Monte- Carlo code which includes the size distribution function [34]. This incorporation of size distribution function enabled us to generate a more realistic scattering profile what we obtain otherwise by performing light scattering experiments. The overall comparative analyses of the scattering profiles suggest that although the profiles follows same trends, there is a need to understand the azimuthal dependency of scattering elements for such bio-particles.

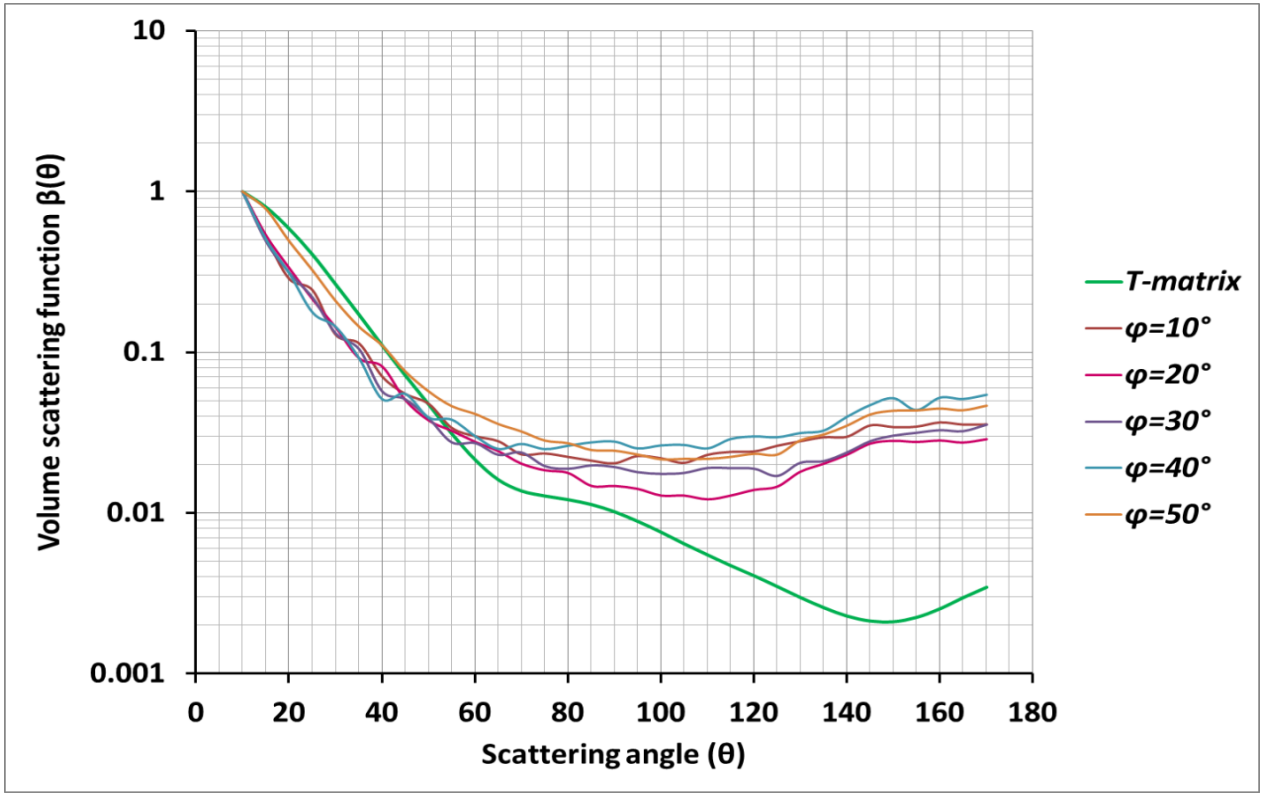

**Figure 3:** Volume scattering function β(θ) for *P. aeruginosa* at 543.5 nm

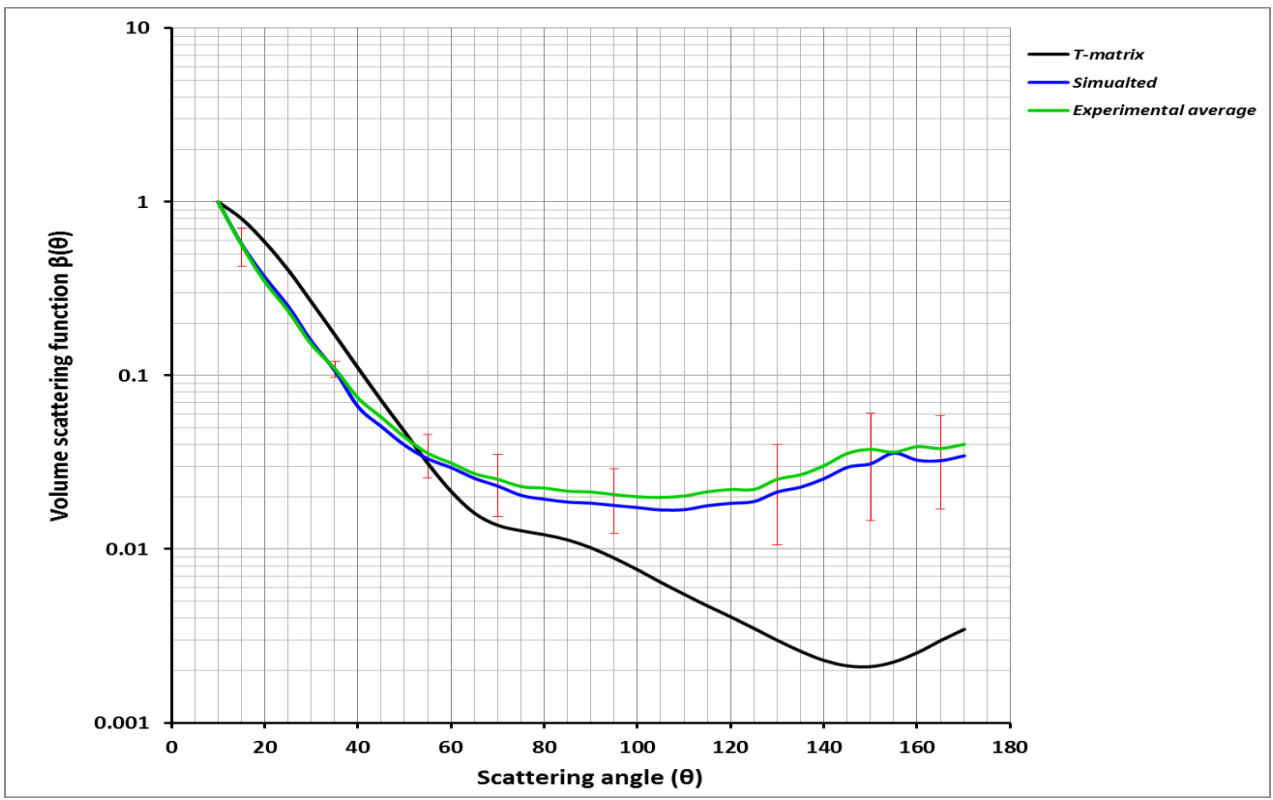

**Figure 4:** Comparative Volume scattering function β(θ) for *P. aeruginosa* at 543.5 nm

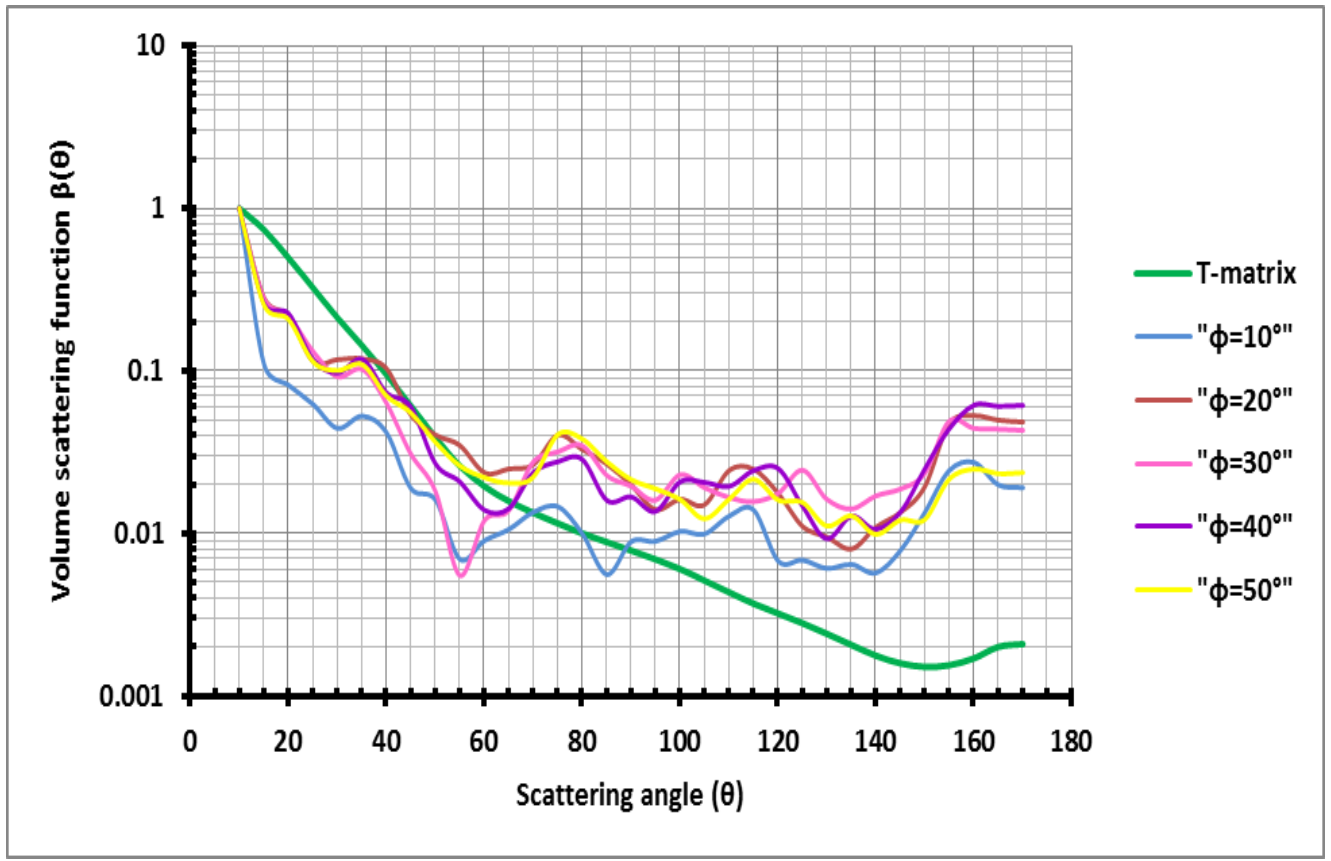

**Figure 5:** Volume scattering function β(θ) for *M. smegmatis* at 543.5 nm

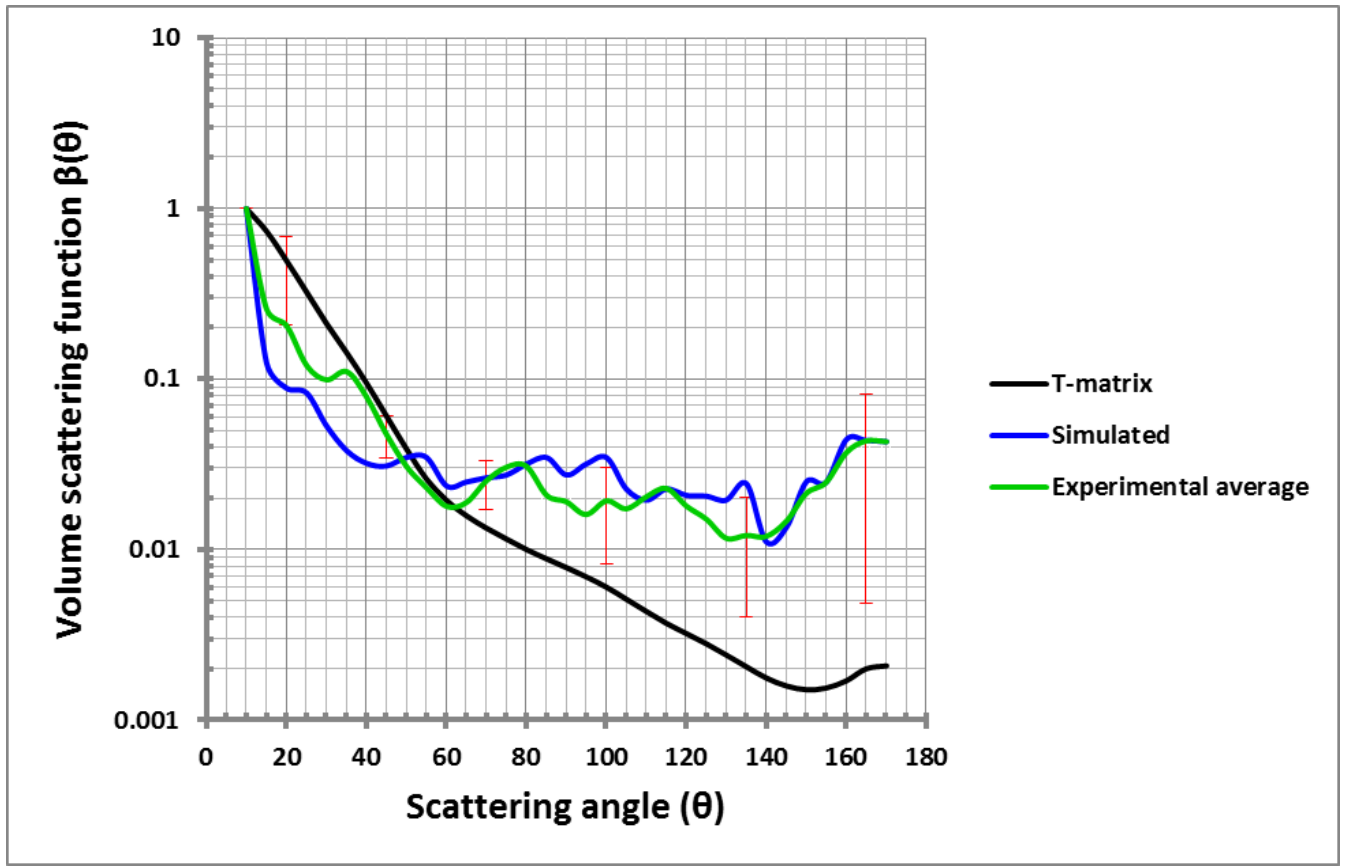

**Figure 6:** Comparative Volume scattering function β(θ) for *M. smegmatis* at 543.5 nm

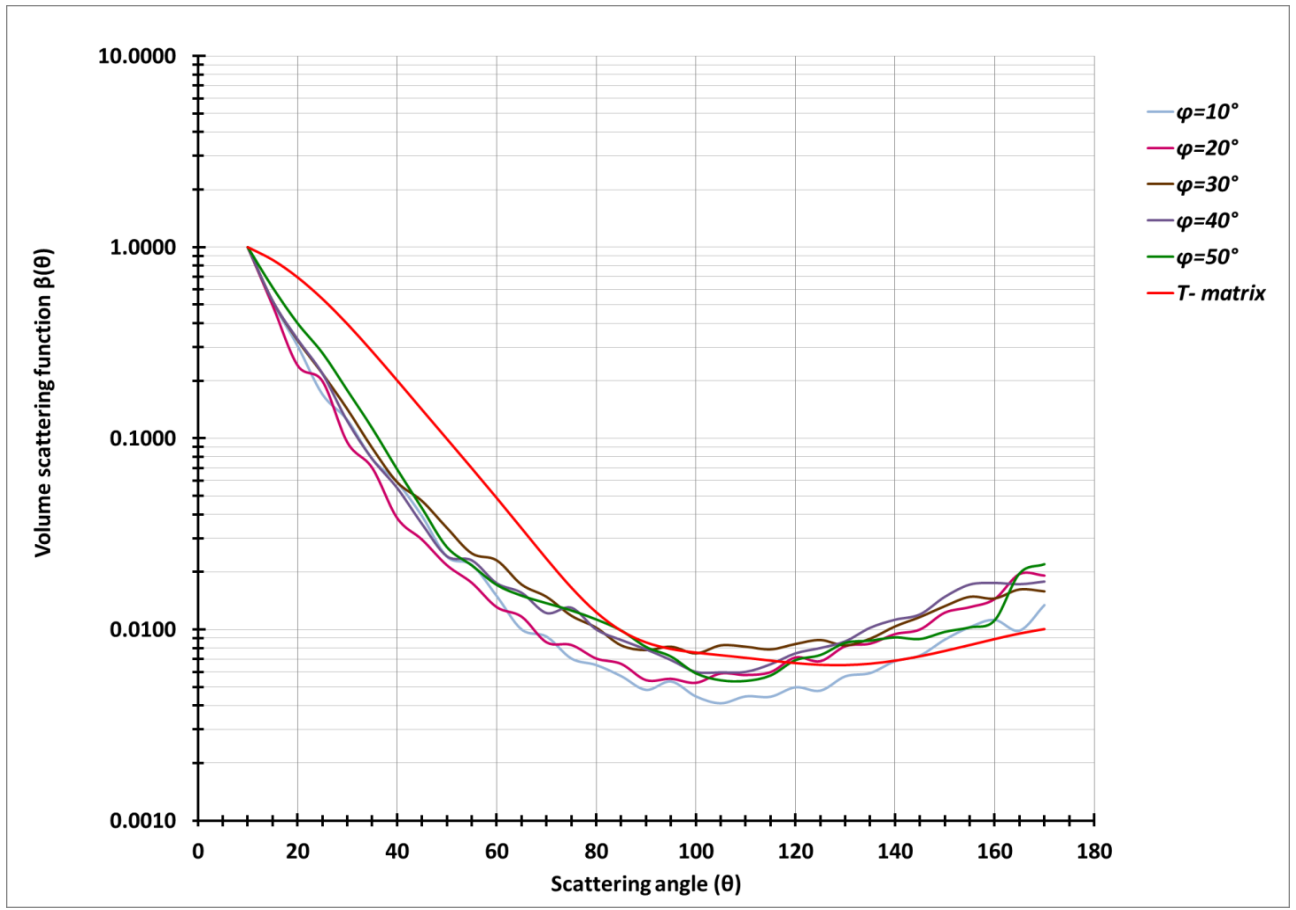

**Figure 7:** Volume scattering function β(θ) for *P. aeruginosa* at 632.8 nm

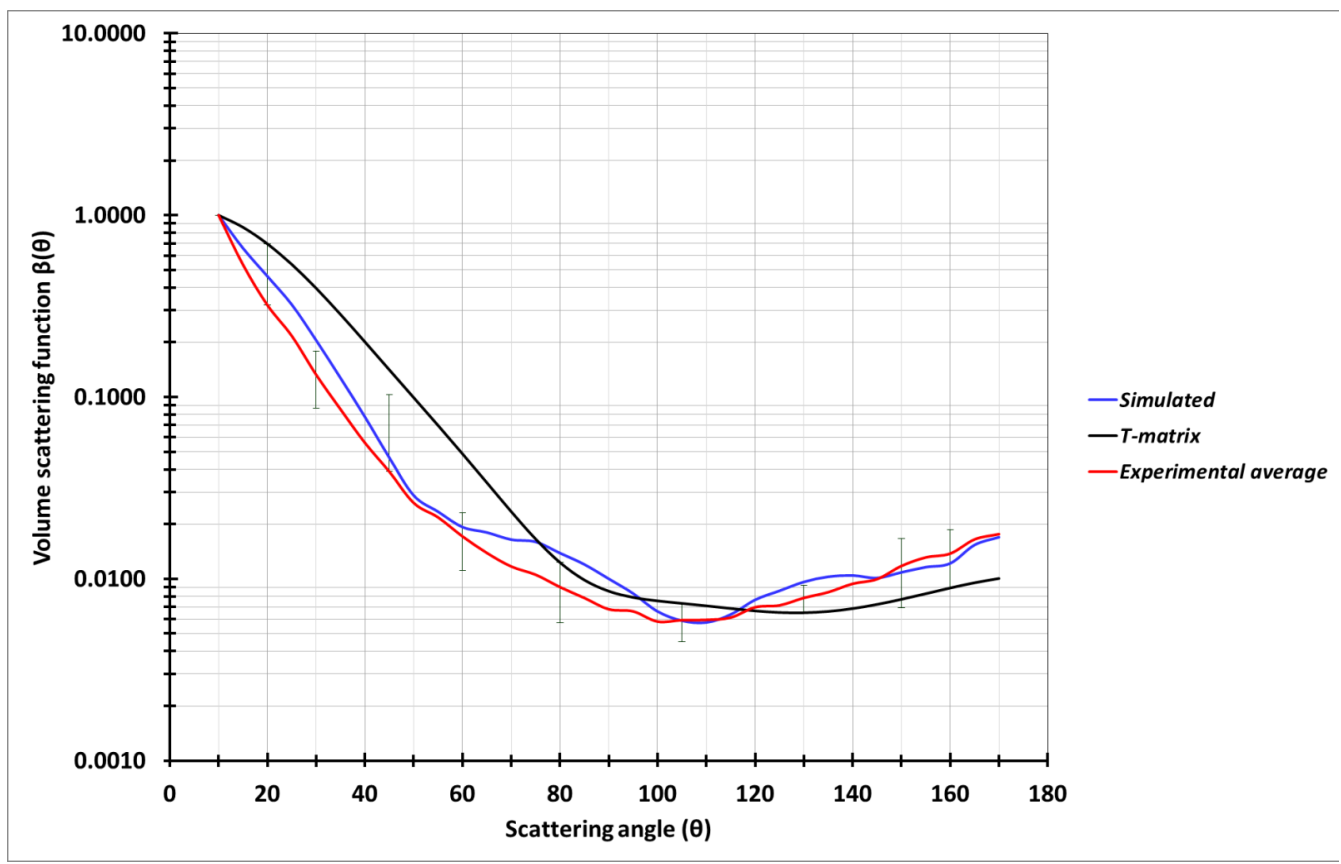

**Figure 8:** Comparative Volume scattering function β(θ) for *P. aeruginosa* at 632.8 nm

## 4. Discussions

The size-dependent scattering profiles of both *M. smegmatis* and *P. aeruginosa* is presented in this paper. We analyzed the refined differences in the characteristic profiles which exhibited azimuthal dependency for both the species. Some small morphological changes in the bio-particles under the influence of any external factors (like metabolic change or drug resistance etc.) can be monitored by light scattering tool. In simple words, one might be able to detect precisely about the morphological changes which occur in bio-particles like bacteria as a consequence of external alterations which otherwise may not be possible by bio- chemical or invasive techniques. The scattering results revealed the unwaveringness and nature of light scattered from such particle systems. Our previous report of light scattering studies on similar samples and our current study on the selected samples point out the necessity and importance of azimuthal dependency while carrying out light scattering investigations [21,34,35]. By carrying out investigations on variety of other bio-particle samples, we can ascertain whether other bioparticles or particles of other origin) also show azimuthal dependency. This also calls for proper choice of the existing popular light scattering theories like Mie-theory and the T-matrix approach. The selection of an appropriate approximation theory depends on the precise characteristics of the particles and also on the extent of errors in the acceptable limit.

**6. Conclusion**

Size quantification of the bio-particles using T-matrix method was carried out. First two significant element of the Mueller matrix was determined successfully. Out work reports the use of azimuthal dependency of bioparticles in light scattering investigations. The azimuthal dependency indicates that although the SEM image reveals random orientation of the particles, they may however get aligned due to some external unknown factors. Investigations on finding clue to these factors is an important and inevitable part of research, particularly while performing experimentation with similar bio-particles. Most importantly, we shall give special effort to calculate the $S_{34}$ element of the Mueller matrix which has drawn much attention in research field other than the other elements of the matrix. Our result may prove to be a vital record for comparative study by other researchers by including the azimuthal component. Finally, our portable light scattering instrument proved to be quite efficient for performing light scattering experiments on different types of small particles including biological particles having exhibited significant profiles.

**References:**

1. M. I. Mishchenko, J. W. Hovenier and Travis, L. D. *Light scattering by nonspherical particles: Theory, Measurements, and Applications*, (Academic Press, San Diego, California 2000).
2. D. Tzarouchis and A. Sihvola, Appl. Sci., **8,** 2, p184, (2018).
3. A. Doicu and M. I. Mishchenko, J. Quant. Spectrosc. Radiat. Transfer , *2019*, https://doi.org/10.1016/j.jqsrt.2019.07.007
4. O. Muñoz, , et al., J. Quant. Spectrosc. Radiat. Transfer , **111**, p187-196, (2010).
5. F. Kuik, P.Stammes and J.W. Hovenier, Appl. Opt., **30,**33, p 4872 – 4881, (1991).
6. H. Volten, , et al. Laboratory Measurements and T-Matrix Calculations of the Scattering

Matrix of Rutile Particles in Water, *Appl. Opt.*, **38**, 24, p5232-5240, (1999).

7. S. Sharma, and D. J. Somerford, In *Light scattering by optically soft particles: Theory and applications*, (Praxis Publishing, Chichester, UK, 2006).
8. C. Saunders et al., Ann. Geophysics, **16**, 618-627, (1998).
9. Y. Takano, et al. Solar radiative transfer in Cirrus clouds. Part I: Single scattering and optical properties of hexagonal ice crystals, J. Atmosph. Sci., **46**, 1, p 3-19, (1989).
10. M. S. Quinby-Hunt, et al. Polarized-light scattering studies of marine *chlorella*. *Limnol. Oceanogr.*, **34**, 8, p1587–1600, (1989).
11. H. Volten, et al. Laboratory measurements of angular distributions of light scattered by phytoplankton and silt, Limnol. Oceanogr., **43**, 6, p1180-1197, (1998).
12. D. Daugeron et al., Meas. Sci. Technol., 18, p632–638, (2007).
13. A. Ben-David, Appl. Opt., **38**, p2616-2624, (1999).
14. N. N. Boustany, S. A. Boppart, and V. Backman, *Annual Review of Biomedical Engineering*, https://doi.org/10.1146/annurev-bioeng-061008-124811, (2010).
15. A. Katz et al., IEEE, **9**, 2, p 277-287, (2003).
16. L. Kolokolova, W. Sparks and D Mackowski, In *Astrobiological remote sensing with circular polarization, Polarimetric Detection, Characterization and Remote Sensing*, (Springer, Dordrecht; p 277-294, doi:10.1007/978-94-007-1636-0_11, 2011).
17. A. N. Shvalov et al., Cytometry, 41, 1, p41-50, (2000).
18. Z. Hu, *et al.*, Nanoscale, **41**, p 19233- 19642, (2018).
19. B. J. Berne, Accounts of Chemical Research, **6**, 9, p318-322, (1973).

20. S. Roy, et al., Monitoring of pathogen carrying air-borne tea dust particlesby light scattering, J. Quant. Spectrosc. Radiat. Transfer, **112**, p1784- 1791, (2011).
21. S. Roy et al., Study of ZnO nanoparticles: Antibacterial property and light depolarization property using light scattering tool**,** J. Quant. Spectrosc. Radiat. Transfer, **118**, p 8-13, (2013).
22. P. J.Wyatt, Appl Opt ; **7**: p 1879–1896, (1968).
23. A. Diaspro, G. Radicchi, C. Nicolini, IEEE Trans Biomed Eng., **42**, p1038–1043, (1995).
24. V. P. Maltsev, K. A.Semyanov, Characterisation of Bio-Particles from Light Scattering. Utrecht: VSP ,132p, (2004).
25. *K. Rauf, et al. Study of putative microfossils in space dust from the stratosphere,* Int. J. Astrobiol.*,* **9**, 3, p183-189, (2010).
26. C. Wickramasinghe, Int. J. Astrobiol., **10**, p 25-30, (2011).
27. National Nosocomial Infections Surveillance (NNIS). System Report, data summary from January 1992 through June 2004, issued October 2004. Am. J. Infect. Control. **32**, 8, p 470–485, (2004).
28. C.A. Kimberly, and M. C. Wolfgang, Current issues in molecular biology, **14** , 2,p 47, (2012).
29. A. P. Magiorakos, A. Srinivasan, R. B. Carey et al. Clin Microbiol Infect , **18**, p 268–81, (2012).
30. G. King, Applied and Environmental Microbiology, **69**, p 7266–7272, (2003).
31. H. Alderton and D. Smith , In: *Safety in the laboratory*, editors: Parish T, Stoker NG, ( Humana Press; p. 367–383, 2001).
32. M. U. Shiloh and P. A. *,* Di Giuseppe Champion Curr Opin Microbiol., **13**, 1,p 86–92, (2010).
33. Quintby- Hunt, and A. J. Hunt, *In Proceedings of SPIE*, **925**, p 288-295, (1988).
34. S. Roy and G. A. Ahmed, Optik-Int. J. Light Electron Opt. , **122**, p 1000-1004, (2011).
35. S. Roy, as discussed in https://www.giss.nasa.gov/staff/mmishchenko/ELS-XVI/Contributed/Roy_1.pdf, (2018).

**Declaration of interests**

☑ The authors declare that they have no known competing financial interests or personal relationships that could have appeared to influence the work reported in this paper.

☐ The authors declare the following financial interests/personal relationships which may be considered as potential competing interests:

**Sanchita Roy :** Conceptualization, Validation, Data curation ,Writing- Original draft preparation, Writing- Reviewing and Editing

**Nilakshi Barua:** Methodology, Resources, Investigation, Data curation.

**Aranya Bhuti Bhattacherjee***:* Visualization, Data curation.

*Gazi A Ahmed*: Supervision, Validation, Software.